# Negative Refractive Index in Artificial Metamaterials

A. N. Grigorenko

*Department of Physics and Astronomy, University of Manchester, Manchester, M13 9PL, UK*

We discuss optical constants in artificial metamaterials showing negative magnetic permeability and electric permittivity and suggest a simple formula for the refractive index of a general optical medium. Using effective field theory, we calculate effective permeability and the refractive index of nanofabricated media composed of pairs of identical gold nano-pillars with magnetic response in the visible spectrum.

The refractive index of an optical medium, $n$, can be found from the relation $n^2 = \varepsilon\mu$, where $\varepsilon$ is medium's electric permittivity and $\mu$ is magnetic permeability.[1] There are two branches of the square root producing $n$ of different signs, but only one of these branches is actually permitted by causality.[2] It was conventionally assumed that this branch coincides with the principal square root $n = \sqrt{\varepsilon\mu}$.[1,3] However, in 1968 Veselago[4] suggested that there are materials in which the causal refractive index may be given by another branch of the root $n = -\sqrt{\varepsilon\mu}$. These materials, referred to as left-handed (LHM) or negative index materials, possess unique electromagnetic properties and promise novel optical devices, including a perfect lens.[4-6] The interest in LHM moved from theory to practice and attracted a great deal of attention after the first experimental realization of LHM by Smith *et al.*[7], which was based on artificial metallic structures proposed by Pendry *et al.*[8,9] Recently, the working frequency of negative index materials has been extended to visible[10] and infrared light.[11]

The choice of the causal branch of the refractive index becomes therefore of practical importance and has been addressed[12], where a rather complicated procedure for the branch choice has been suggested, see also review.[6] The aim of this letter is to suggest a unique and simple analytical expression that gives refractive index of an optical medium and apply it to calculate optical constants of the recent nanomaterials with magnetic response in visible spectrum[10] within effective field theory.

Let us consider a plane electromagnetic wave propagating in a homogeneous and isotropic medium. Such a wave is conventionally described by an exponential factor $\exp(i\frac{n\omega}{c}x - i\omega t)$, where $\omega$ is the light angular frequency, $x$ is the axis of propagation, and $c$ is the speed of light.[1] The loss of energy (or a gain of energy in an active medium) per second per unit volume is given by an expression[2] $Q = \frac{\omega}{8\pi} \left( \varepsilon'' |E|^2 + \mu'' |H|^2 \right)$, where the $\varepsilon'' = \mathrm{Im}(\varepsilon)$, $\mu'' = \mathrm{Im}(\mu)$, $E$ and $H$ are the complex amplitudes of the electric and magnetic field respectively. It is clear that a wave should decay in the direction of its propagation (and hence we should choose $\mathrm{Im}(n) > 0$) in an optical medium with positive losses ($Q > 0$), while a wave should be amplified in the direction of its propagation (and hence we should choose $\mathrm{Im}(n) < 0$) in an active optical medium with gain ($Q < 0$). Therefore, the signs of $Q$ and imaginary part of $n$ coincide, which gives us a unique way to choose the causal value of the refractive index. For a plane wave[2] $\varepsilon E^2 = \mu H^2$ and we can rewrite the loss (gain) as $Q = \frac{\omega}{8\pi} |\varepsilon| |E|^2 \left( \frac{\varepsilon''}{|\varepsilon|} + \frac{\mu''}{|\mu|} \right)$ or $Q = \frac{\omega}{4\pi} |\varepsilon| |E|^2 \sin\left( \frac{\arg(\varepsilon) + \arg(\mu)}{2} \right) \cos\left( \frac{\arg(\varepsilon) - \arg(\mu)}{2} \right)$, where $\arg(z)$ is the argument of $z$. This implies that signs of $Q$ and $\mathrm{Im}(n)$ will coincide if we choose the refractive index as

$$n = {}_{cas}\sqrt{\varepsilon\mu} \equiv \sqrt{|\varepsilon\mu|} \cdot \exp\left( i \frac{\arg(\varepsilon) + \arg(\mu)}{2} \right) \mathrm{Sign}\left[ \cos\left( \frac{\arg(\varepsilon) - \arg(\mu)}{2} \right) \right], \quad (1)$$

where $\mathrm{Sign}[x]$ gives the sign of $x$ and $\sqrt{\phantom{x}}$ is the principal square root. Using simple algebra, we can rewrite the causal index of refraction as

$$n = \sqrt{\varepsilon}\sqrt{\mu} \cdot \mathrm{Sign}\left[ \mathrm{Re}\left( \frac{\sqrt{\varepsilon}}{\sqrt{\mu}} \right) \right], \quad (2)$$

where $\sqrt{\phantom{x}}$ is the principal square root. To a certain extent, the expression (1) (or (2)) is a trivial consequence of causality. However, to the best of our knowledge, this simple formula is not mentioned in LHM literature.[6] (Often, as was suggested by Referee, researchers use an expression $n = \sqrt{\varepsilon\mu} \cdot \text{Sign}\left[\text{Im}(\sqrt{\varepsilon\mu})\right]$. For any medium this expression should be rewritten as $n = \sqrt{\varepsilon\mu} \cdot \text{Sign}\left[\text{Im}(\sqrt{\varepsilon\mu})\right] \cdot \text{Sign}\left[\frac{\varepsilon"}{|\varepsilon|} + \frac{\mu"}{|\mu|}\right]$ ).

The case of zero losses $Q$=0 requires special consideration. Zero losses occur in a medium whose optical constants satisfy either $\frac{\varepsilon}{|\varepsilon|} = -\frac{\mu}{|\mu|}$ or $\frac{\varepsilon}{|\varepsilon|} = \frac{\mu^*}{|\mu|}$. In the former case, $\cos\left(\frac{\arg(\varepsilon) - \arg(\mu)}{2}\right) = 0$ and the sign of cosine used in the formula (1) is not defined. The imaginary part of $n$ is not zero in this case (provided $\text{Re}(\varepsilon) \neq 0$) and therefore waves could either decay or be amplified in the direction of propagation. Since $Q$=0 we should choose the decaying waves and $n$ with $\text{Im}(n) > 0$. In the latter case of $\frac{\varepsilon}{|\varepsilon|} = \frac{\mu^*}{|\mu|}$, the refractive index is real and we should choose its sign in accordance with the original Veselago's suggestion[4] (positive when $\text{Re}(\varepsilon) > 0$ (hence $\text{Re}(\mu) > 0$) and negative when $\text{Re}(\varepsilon) < 0$ (hence $\text{Re}(\mu) < 0$)). Finally, there exists a degenerate case where both conditions mentioned above are met so that $\text{Re}(\varepsilon)=0$, $\text{Re}(\mu)=0$, and $\text{Sign}(\text{Im}(\varepsilon))=-\text{Sign}(\text{Im}(\mu))$ (e.g., $\varepsilon=-i$ and $\mu=i$). We believe that such a medium is inherently birefringent and show both positive and negative values of the (real) refractive index.

Obviously, the causal refractive index (1), (2) should be used to characterise LHM. To be specific, let us consider an optical medium in which the electric response is generated by an "electric" resonant mode of constituent "molecules" contributing to $\varepsilon$ of a dilute LHM as[1] $\varepsilon(\lambda) = 1 + F_e \lambda^2 / (\lambda^2 - \lambda_e^2 - i\lambda\Delta\lambda_e)$, where $\lambda_e$ is the wavelength of the electric resonance, $\Delta\lambda_e$ its half-width and $F_e$ the effective oscillator strength, while the magnetic response is generated by another resonant mode of "molecules" contributing to $\mu$ by the "Pendry-type" expression[8] $\mu(\lambda) = 1 + F_m \lambda_m^2 / (\lambda^2 - \lambda_m^2 - i\lambda\Delta\lambda_m)$, where $\lambda_m$, $\Delta\lambda_m$ and $F_m$ have the same meaning as above but for the magnetic resonance. Figure 1(a) shows the spectral dependence of $n$ of such LHM calculated using the principal square root $n = \sqrt{\varepsilon\mu} = n' + in''$. One can see that the principal branch does not adequately describe the spectral behaviour of $n$ as it yields the negative sign of the imaginary part $n''$ in the region 0.3-0.6μm (and wrong positive $n'$) which contradicts causality. Figure 1(b) shows the spectral dependence of the causal refractive index (1) ($n = {}_{cas}\sqrt{\varepsilon\mu}$) that yields the correct $n$ in the whole spectral range including the region 0.3-0.6μm with negative $n'$.

The choice of $n$ has a dramatic effect on the effective optical constants obtained within effective medium theory, where the interaction between LHM "molecules" is not weak and affects the resonant properties of individual "molecules" (e.g., changes the resonant wavelengths, $\lambda_{e,m}$, half-widths, $\Delta\lambda_{e,m}$, etc.). We illustrate this by calculating the effective permeability for a dense LHM made of the same "molecules". According to effective medium theory, the effective field acting on a "molecule" in a dense material is given by the Lorentz-Lorenz expression[1] (in the limit $s \ll a \ll \lambda$, where $s$ is the size of the "molecule", $a$ is the average distance between "molecules" and $\lambda$ is light wavelength). Also, the effective resonant parameters of the "molecules" (effective $\lambda_{e,m}^{eff}$ and $\Delta\lambda_{e,m}^{eff}$) in a dense LHM become functions of effective $\varepsilon_{eff}$ and $\mu_{eff}$. In a first approximation we assume that the shift of the resonant wavelengths induced by the neighbouring "molecules" is proportional to $n_{eff}$: $\delta\lambda_{e,m}^{eff} = q_{e,m}n_{eff}$, where $q_{e,m}$ are small constants ($q_{e,m} \ll \lambda_{e,m}^{eff}$). This approximation is supported by[13-15], where the resonant wavelengths of metallic "molecules" are shown to be proportional to the refractive index of the environment. To find the effective permeability we therefore solve the pair of self-consistent Lorentz-Lorenz (Clausius-Mossotti) equation.[1] Figures 1(c) and (d) show the effective permeability $\mu_{eff}$ calculated in the spectral range near the "magnetic resonance" using the principal branch $n_{eff} = \sqrt{\varepsilon_{eff}\mu_{eff}}$ and the causal branch $n_{eff} = {}_{cas}\sqrt{\varepsilon_{eff}\mu_{eff}}$ of the refractive index, respectively, ($q_{e,m}$=4.5nm). It is clear that Figs. 1(c) and (d) give completely different dependences for $\mu_{eff}$ near the resonance.

Finally, we apply the causal effective medium theory to practice. We have recently fabricated artificial nanomaterials formed by regular arrays of "nano-molecules" produced by pairs of electromagnetically coupled identical gold nano-pillars with plasmon resonances in the visible part of the spectrum.[10] Figure 2(a) shows an electron micrograph of one of our samples. The prepared structures were regular arrays of Au pillars fabricated by high-resolution electron-beam lithography on a glass substrate and grouped in tightly spaced pairs. At small separation, near-field coupling between neighbouring pillars within a pair is essential and plasmon resonance observed for an individual pillar splits into two resonances for a pillar pair. These resonances are referred to as symmetric and antisymmetric. For the symmetric resonance, electrons in neighbouring pillars move in phase and generate an overall dipole contribution to $\varepsilon$. In the antisymmetric mode, however, the electrons move in anti-phase so that the oscillating dipoles cancel each other, leaving only the overall magnetic response contributing to $\mu$ and/or quadrupole response contributing to non-diagonal, non-local $\varepsilon$. Figures 2(b) and (c) show the calculated current distributions for the symmetric and antisymmetric $z$-modes, respectively. The symmetric $z$-mode contributes to $\varepsilon_z$. The overall dipole moment of the antisymmetric $z$-mode is zero and the circulating currents in the $x$-$z$ plane produce $\mu_y$.

Figures 2(d) and (e) show typical reflection spectra measured on the sample of Fig. 2(a) under conditions of normal light incidence for TM light (with the electric field vector along the *x*-axis) and TE light (with the electric field vector along the *y*-axis), respectively. There are two distinct resonance peaks in the TM spectrum indicated by arrows and only one peak in the TE spectrum. The symmetry analysis and the numerical solution of Maxwell equations[10] (shown in the insets of Fig. 2(d) and (e)) prove that the weaker resonance peak (observed at green wavelengths) corresponds to the antisymmetric *z*-mode and the stronger peaks of Fig. 2(d) and (e) (observed at red wavelengths) correspond to the symmetric *x*- and *y*-modes of plasmonic resonances, respectively. Our best samples (covered with a thin glycerine layer)[10,13] show negative index of refraction *n*=-0.7 at green light, with the quality ratio Re(*n*)/Im(*n*)=0.4.

It turned out that the reflection spectra shown in Fig. 2(d) and (e) are described extremely well by the Fresnel coefficients of a thin film placed on a glass substrate[1] with film's $\varepsilon$ and $\mu$ given by the standard dispersion relations described above.[10] (Optical thickness of the film was calculated using the causal refractive index *n*). Such behaviour is not surprising for a dilute LHM where the interaction between pillar pairs is weak and dispersion of an individual "molecule" shapes the spectral dependence of $\varepsilon$ and $\mu$. However, it is unusual for dense LHM.[16] The causal effective mean theory resolves this contradiction. Figure 3 presents the magnetic permeability and refractive index obtained within the causal effective field theory for the nanofabricated material of Fig. 2. In these calculations we expressed the effective Lorentz-Lorenz field as a sum over the 2D periodic array[17] and simultaneously solved self-consistent Clausius-Mossotti equations for $\varepsilon$ and $\mu$ (with demagnetization and depolarization factors corresponding to the 2D array). The solid line of Fig. 3 shows the calculated permeability for the experimental array of Fig. 2, the long-dashed line gives $\mu$ for the array with a double density of pillar pairs and the short-dashed line presents $\mu$ for the array with a twice-smaller density of pairs. It is clear that the dispersion of $\mu$ in all 3 cases is described well by the standard dispersion relation, which explains the success of the Fresnel coefficients in modelling the reflection spectra from the fabricated 2D arrays.

In conclusion, we suggested a simple analytic expression for the refractive index of a general optical medium and applied it to calculate optical constants of double-pillar arrays within effective field theory.

The support of Paul Instrument Fund is acknowledged.

**Figure captions.**

Fig. 1. Refractive index calculated with (a) the principal square root and (b) the causal square root. Effective permeability calculated within effective field theory with (c) the principal square root and (d) the causal square root. LHM parameters are: $\lambda_e$ =1µm, $\Delta\lambda_e$ =0.1µm, $F_e$=10 and $\lambda_m$=0.5µm, $\Delta\lambda_m$ =0.04µm, $F_m$=0.22.

Fig. 2. Nanofabricated medium with magnetic response in the visible spectrum. (a) A micrograph of the sample. (b) The distribution of electric currents (conical arrows) inside a pair of pillars for the resonant symmetric $z$-mode. (c) Same for the antisymmetric $z$-mode. (d, e) Experimental reflection spectra measured for TM and TE polarizations, respectively (solid lines). The insets show the current distribution calculated by solving Maxwell equations for the actual experimental geometry at the resonant wavelengths and the reflection calculated with Fresnel coefficients (squares). The resonance parameters are: (d) $\lambda_e$ =0.69µm, $\Delta\lambda_e$ =0.23µm, $F_e$=3.9 and $\lambda_m$=0.55µm, $\Delta\lambda_m$=0.082µm, $F_m$=0.1, (e) $\lambda_e$ =0.64µm, $\Delta\lambda_e$ =0.18µm, $F_e$=3.

Fig. 3. Calculated dispersion of the real part of (a) magnetic permeability and (b) the index of refraction in the array of Fig. 3 within effective field theory. (1) $a$=707nm, (2) $a$=500nm, (3) $a$=353nm.

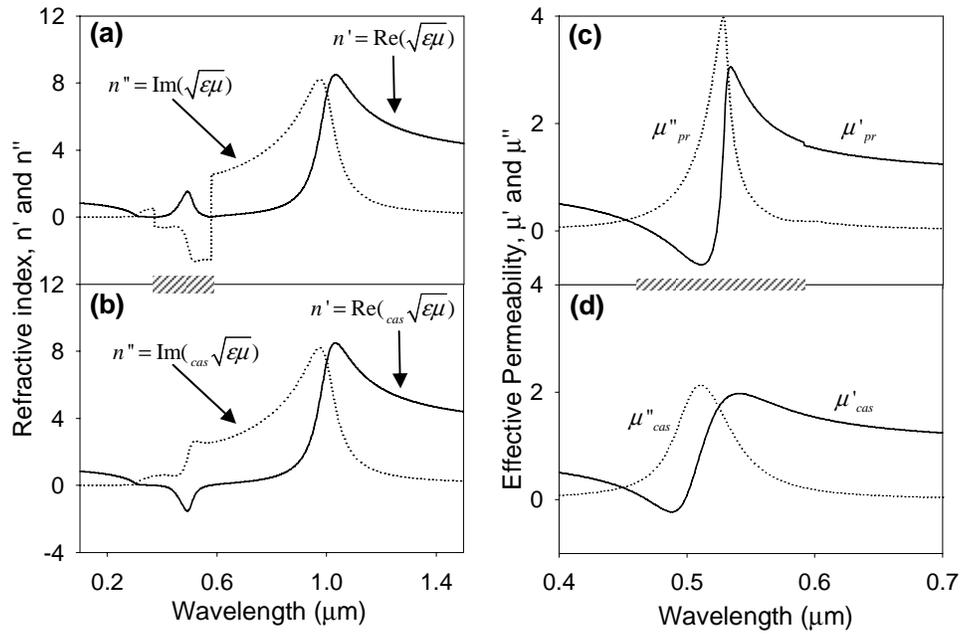

Fig. 1.

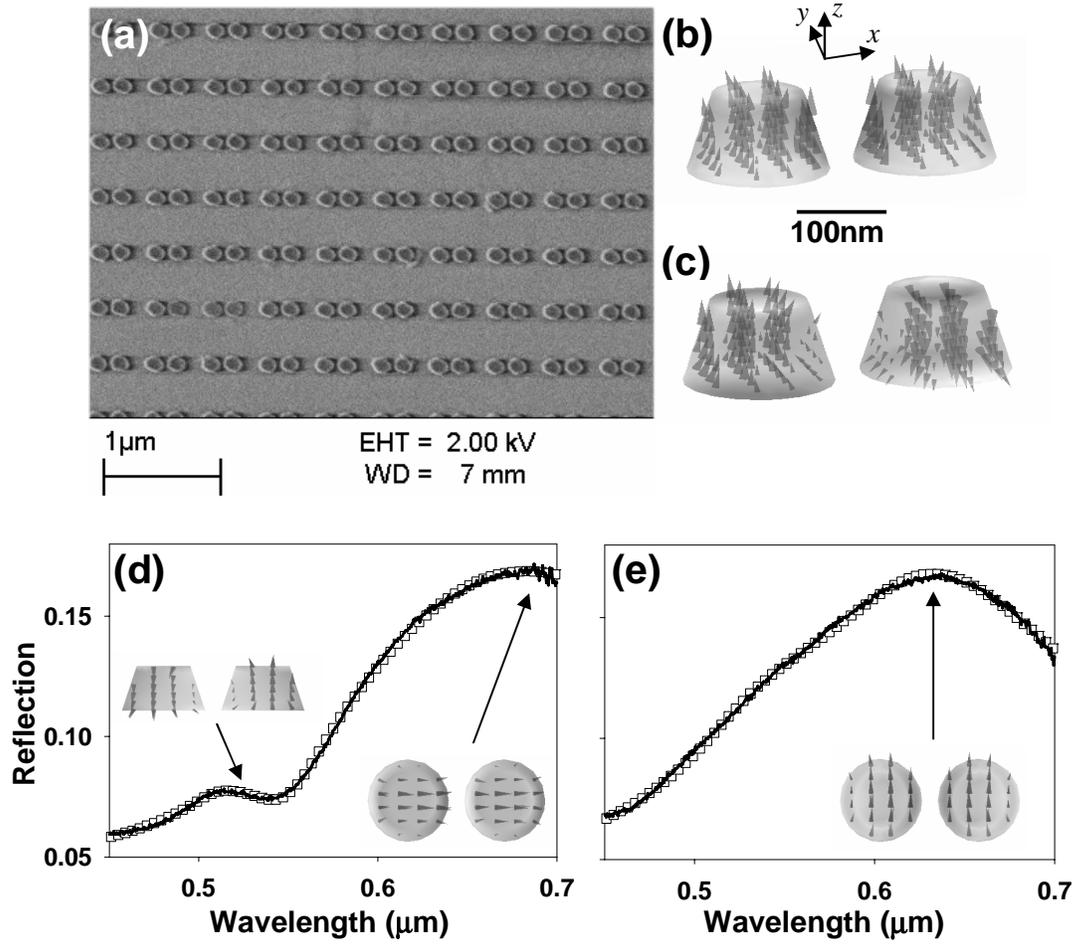

Fig. 2.

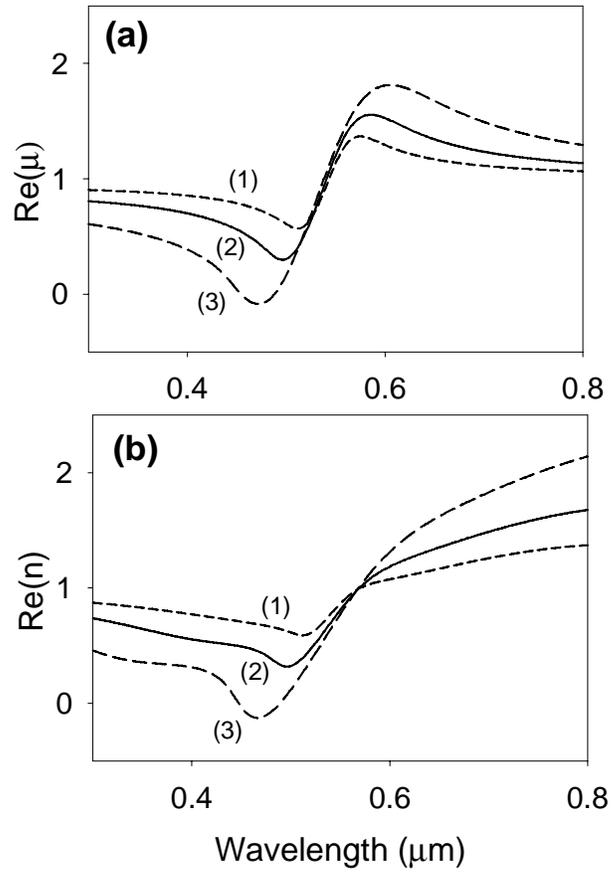

Fig. 3.